\documentclass[pra, preprint, a4paper,showpacs,superscriptaddress]{revtex4-1}
\usepackage{amsmath,amscd,amsfonts,amssymb, amsbsy, color,bbm, bm}
\usepackage{enumitem}
\usepackage{braket}
\usepackage{graphicx}
\usepackage{dcolumn}

\numberwithin{subsection}{section}
\numberwithin{equation}{section}
\usepackage{natbib}

\begin{document}

\title{Quantum hypothesis testing and state discrimination}
\author{Prabhu Tej}\email{prabhutej.j@gmail.com}
\affiliation{Department of Physics, Bangalore University,
	Jnana Bharathi, Bengaluru, Karnataka 560056.}
\author{Syed Raunaq Ahmed}\email{thebeatofadifferentdrum@gmail.com}
\affiliation{Department of Physics, Bangalore University,
	Jnana Bharathi, Bengaluru, Karnataka 560056.}
\author{A.R.Usha Devi}\email{arutth@rediffmail.com}
\affiliation{Department of Physics, Bangalore University,
	Jnana Bharathi, Bengaluru, Karnataka 560056.}
\affiliation{Inspire Institute Inc., Alexandria, Virginia, 22303, USA.}
\author{A. K. Rajagopal}\email{attipat.rajagopal@gmail.com} 
\affiliation{Inspire Institute Inc., Alexandria, Virginia, 22303, USA.}

\date{\today}

\begin{abstract}
This expository article gives an overview of the theory of hypothesis testing of quantum states in finite dimensional Hilbert spaces. Optimal measurement strategy for testing binary quantum hypotheses, which result in  {\em minimum error probability}, is discussed. Collective and individual adaptive measurement strategies in testing hypotheses in the multiple copy scenario, with various upper and lower bounds on error probability, are outlined. A brief account on {\em quantum channel discrimination} and the role of entangled states in achieving {\em enhanced precision} in the task of channel discrimination is given.    
\end{abstract}

\maketitle

\section{\label{sec:level1}Introduction}
Given two quantum states $\rho_0$ and $\rho_1$, estimating the true state, based on an optimal decision strategy, in favour of one of the binary hypotheses $H_0$ or $H_1$ is referred to as {\em quantum (binary) hypothesis testing}. The first step towards the mathematical description of quantum hypothesis testing was formulated by Helstrom~\cite{Helstrom1967, Helstrom}. Further progress in testing quantum hypotheses was made by Yuen, Kennedy and Lax~\cite{Yuen1970, YKL1970}, Holevo~\cite{Holevo}, Parthasarathy~\cite{KRP2001}, Hayashi~\cite{Hayashi}, Kargin~\cite{kargin2005}, Nussbaum and Szkola~\cite{NS2006}, Audenaert et. al.,~\cite{qcb_prl_07, qcb_cmp2008}.    

A quantum system is described by a {\em density operator} $\rho$, which is a non-negative operator in a complex Hilbert space $\mathcal{H}$, with unit trace. A  set consisting of finite number of positive operators $\{E_\alpha\}$ obeying 
\begin{equation} 
\label{measurement}
E_\alpha \geq 0,\ \ \ \ \sum_{\alpha}E_{\alpha}= I, 
\end{equation}
characterize measurement with a countable number of outcomes $\alpha=0,1,2,\ldots, d$. This set is referred to as {\em positive operator valued measure} (POVM))~\cite{nc2000}.  Every element $E_\alpha$ of the POVM corresponds to a measurement outcome $\alpha$. Measurement in a quantum state  $\rho$ results in an outcome $\alpha$ with probability     
\begin{equation} 
\label{prob}
p_\alpha={\rm Tr}(\rho\, E_\alpha).  
\end{equation}
In this article we confine our discussion only to finite dimensional complex Hilbert spaces.       

In binary hypothesis testing,  the problem is to decide,  which of the two density matrices $\rho_0$ and $\rho_1$ is  true, based on  a measurement strategy leading to   minimum probability of error. Suppose the hypotheses $H_0$, $H_1$ are given by quantum states $\rho_0$ and $\rho_1$, with respective prior probabilities $\Pi_0$ and $\Pi_1;$ $\Pi_0+\Pi_1=1$. Then, probabilities of making incorrect decision are given by  
\begin{equation} 
\label{probability}
p(\beta | H_\alpha)={\rm Tr}(\rho_{\alpha}\,E_{\beta}), \alpha\neq \beta=0,1.  
\end{equation}
Type I error $p(1\vert H_0)={\rm Tr}(\rho_{0}\,E_{1})$ is the error of accepting the {\em alternative hypothesis} $H_1$, when the {\em null hypothesis} is  true. Type II error $p(0\vert H_1)={\rm Tr}(\rho_{1}\,E_{0})$ occurs when {\em alternative hypothesis} $H_1$ is the true one in reality, but  {\em null hypothesis} is accepted. 
  
An optimal decision strategy requires one to recognize a measurement POVM $\{E^{\rm opt}_\alpha,\, \alpha~=~0,1\},$ such that the {\em average probability of error} 
\begin{eqnarray}
P_e=\Pi_0\, p(1|H_0)+ \Pi_1\, p(1|H_0)=\Pi_0\,{\rm Tr}(\rho_{0}\,E_1)+\Pi_1\,{\rm Tr}(\rho_{1}\,E_0)    
\end{eqnarray}  
is minimum. It may be noted that when $\rho_0$ and $\rho_1$ commute with each other, the problem reduces to the testing of hypotheses based on classical statistical decision strategy. The optimal decision in the classical hypothesis test is realized by the {\em maximum-likelihood decision rule}~\cite{Helstrom}.  

In the case when null hypothesis $H_0$ is assigned to $\rho_0^{\otimes M}$ (i.e.,  tensor product of $M$ copies of the state $\rho_0$), and the alternative hypothesis to the tensor product  $\rho_1^{\otimes M}$,  the {\em asymptotic error rate}, realized in the limit of $M\rightarrow \infty$, is of interest~\cite{qcb_prl_07, qcb_pra_08}. In the classical setting, the error probability in distinguishing two probability distributions $p_0(\alpha)$ and $p_1(\alpha)$ decreases exponentially with the increase of the number $M$ of statistical trials i.e,  
\begin{equation}
\label{error_exponent}
P_e^{(M)}\sim e^{-M\,\xi(p_0,p_1)}.
\end{equation}
Here, $\xi(p_0,p_1)>0$ denotes the error rate exponent. More specifically, in an optimal hypothesis test, the probability of error $P_e^{(M)}$ decreases exponentially with the increase of the number $M$ of statistical trials.  Chernoff~\cite{chernoff1952} derived the following expression  
\begin{equation}
\label{classical_chernoff}
\xi_{\rm CB}=- \lim_{M\rightarrow\infty}\left(\frac{1}{M}\, \log\, P_{e, {\rm CB}}^{(M)}\right)= 
- \log\,  \inf_{s\in[0,1]}\,\sum_\alpha \left[\, p_{0}^s(\alpha) p_{1}^{1-s}(\alpha)\,\right],
\end{equation}
for the error rate exponent,  which holds exactly in the asymptotic limit of $M\rightarrow \infty$. The error rate exponent $\xi_{\rm CB}$  gives  the asymptotic efficiency of testing classical hypotheses. Moreover, for finite number of trials, one obtains a Chernoff upper bound   $P_{e,\, {\rm CB}}^{(M)}\geq P_{e}^{(M)}$  on  the  probability of error $P_e^{(M)}$. 

A quantum generalization of the Chernoff's result remained unsolved for long time. Various lower and upper bounds on the optimal error exponent in terms of 
fidelity between the two density operators $\rho_0$, $\rho_1$  were identified~\cite{kargin2005}.  Nussbaum
and Szkola~\cite{NS2006}, and Audeneart et. al.~\cite{qcb_pra_08} settled the issue by identifying  the quantum Chernoff bound  
\begin{equation}
\xi_{\rm QCB}=- \lim_{M\rightarrow\infty}\, \left(\,\frac{1}{M}\, \log\, P_{e, {\rm QCB}}^{(M)}\right)= 
-  \log\, \inf_{s\in[0,1]}\,{\rm Tr} \left(\, \rho_0^s\, \rho_{1}^{1-s}\,\right), 
\end{equation}
where $P_{e, {\rm QCB}}^{(M)}$ offers an lower bound on probability of error  $P_{e}^{(M)}$.    

In order to arrive at a decision with  minimum error probability one has to choose  optimal measurements for discriminating the states $\rho_0^{\otimes M}$ and $\rho_1^{\otimes M}$. Different measurement strategies employed have been classified into (i) collective measurements, where a single POVM is employed to distinguish $M$ copies of the states $\rho_0$ and $\rho_1$ and  (ii) individual measurements~\cite{acin_pra_2005} performed on each copy of state. As collective measurements, with large number of copies $M$, are hard to achieve in experimental implementation,  individual measurement strategies are prefered. It has been shown~\cite{acin_pra_2005, global_local_pra_11} that individual adaptive measurements, where a sequence of individual measurements designed such that a measurement on any copy is  optimized based on the outcome obtained in previous measurement on the previous copy of the sequence. Such adaptive individual measurement strategies are shown to result in the same precision as that of the collective strategy~\cite{acin_pra_2005}. 

In this paper, we present an overview of quantum state discrimination based on binary hypothesis testing both in the single copy and the multiple copy scenario.  We illustrate, with the help of an example, an alternate approach termed as {\em unambiguous state discrimination}, which is employed for quantum state discrimination. A discussion on collective and adaptive measurements in the multiple copy situation, with various upper and lower bounds on error probability is given in Sec.~3. In Sec.~4 an overview of quantum channel descrimination and the role of entangled states in enhancing precision in the task of channel discrimination is presented. A brief summary is given in Sec.~5.

\section{Quantum hypothesis testing and state discrimination}

Suppose the hypotheses $H_{\alpha},\ \alpha=0,1$ are assigned to the  quantum states characterized by their density
operators $\rho_{\alpha},\, \alpha=0,1$ respectively and  measurements  $\{E_{\beta},\beta=0,1,2...\}$ 
are employed to identify which is the {\em true} state. Let  $p(\beta | H_\alpha),\ \beta\neq \alpha$ denote  the probability with which the hypothesis $\beta$ is declared to be correct, while in fact $\alpha$ is the true one.
Associating an outcome $\beta$ with the measurement $E_{\beta}$, the probability of error in discriminating the states $\rho_0$, $\rho_1$ is given by (see (\ref{measurement}),(\ref{prob}) ) 
$$p(\beta | H_\alpha) = {\rm Tr}(\rho_{\alpha}\, E_{\beta}),\ \ 
\sum_{\beta=0,1}\, p(\beta | H_\alpha)= 1.$$ 
If, with  optimal measurements, one can achieve
\begin{equation}
 p(\beta | H_\alpha) =\delta_{\alpha\beta}=\left\{
\begin{array}{l} 0,\ \ {\rm if}\ \ \alpha\neq \beta \\ 
1, \ \ {\rm if}\ \ \alpha=\beta,
  \end{array} \right. 
 \end{equation} 
then it is possible to arrive at a {\em correct} decision and discriminate the two quantum states $\rho_0$ and $\rho_1$ with {\em no error}. In the special case of orthogonal quantum states  ${\rm Tr}(\rho_0\rho_1)=0$,  the conditions (\ref{probability}) can be expressed in the form of a $2\times 2$ matrix, 
$$\mathbb{P}=\left(\begin{array}{ll}
    {\rm Tr}(\rho_0\, E_0)        &  {\rm Tr}(\rho_0\, E_1)   \\
    {\rm Tr}(\rho_1\, E_0)        &  {\rm Tr}(\rho_1\, E_0)  
\end{array}\right) 
=
\left(\begin{array}{cc}
1 & 0 \\ 
0 & 1 
\end{array}\right).$$ 
and one concludes that {\em orthogonal quantum states can be discriminated perfectly.}
On the other hand, discrimination of non-orthogonal states can only be done with an error.  In order to illustrate this, we consider an example of two pure non-orthogonal states 
$$\rho_i=|\psi_\alpha \rangle\langle \psi_\alpha|,\ \  \alpha=0,1,$$ 
with $\langle \psi_0|\, \psi_1\rangle\neq 0$. Let $E_0$ and $E_1$ be the  measurement operators used to  discriminate these states. Suppose 
\begin{subequations}
\label{assumption}
\begin{align}
\label{assumption1}
{\rm Tr}(\rho_0E_0)=\langle\psi_0|E_0|\psi_0\rangle&=1 \\
\label{assumption2}
{\rm Tr}(\rho_1E_1)=\langle\psi_1|E_1|\psi_1\rangle&=1.
\end{align}
\end{subequations}
Based on the condition (see (\ref{measurement}))   
$$\sum_{\alpha=0,1}E_{\alpha}= I$$ 
on measurement operators, it is readily seen that  $\langle\psi_0|E_1|\psi_0\rangle=0 \, \Rightarrow \sqrt{E_1}\, \vert\,\psi_0\rangle=0$. Then, 
by expressing $|\psi_1\rangle$ as  
\begin{eqnarray*}
|\psi_1\rangle&=& a|\psi_0\rangle+b|\psi_0^{\bot}\rangle, \\
 \langle \psi_0|\psi_0^{\bot}\rangle&=&0,\ |a|^2+|b|^2=1, 0<|b|< 1\ \ \ \ 
\end{eqnarray*}
one obtains 
\begin{equation}
\sqrt{E_1}|\psi_1\rangle=b\, \sqrt{E_1}\,|\psi_0^{\bot}\rangle.
\end{equation}
This in turn implies that 
$$\langle\psi_1|E_1|\psi_1\rangle=|b|^2\neq 1$$ 
in contradiction with \eqref{assumption2}.

{\em Remark}: For a set of orthogonal states, there exists an optimum measurement scheme leading to perfect discrimination,
i.e. with zero probability of error. It is not possible to achieve perfect discrimination of non-orthogonal states in the {\em single copy} scenario.

In a more general setting of testing multiple hypotheses,  a set of  states $\rho_\alpha$ ($\alpha=0,1, \ldots$) are given with apriori probabilities  $\Pi_\alpha$ and a {\em true} state is to be identified from the set of states, by using an adequate measurement strategy. Define {\em average cost} associated with a given strategy as follows~\cite{Helstrom}:  
\begin{eqnarray} 
\label{cost}
\overline{C}=\sum_{\alpha, \beta}\, \Pi_\alpha\, C_{\alpha\beta}\, 
{\rm Tr}(\rho_\alpha\, E_\beta),\ \ \sum_{\beta}\, E_{\beta}=I,
\end{eqnarray}
where $C_{\alpha\beta}$ denotes the {\em cost} incurred when one arrives at a 
{\em wrong decision} (i.e., reaching a conclusion that $\rho_\beta$ is the {\em true} state when, in fact, $\rho_{\alpha}$ happens to be the correct one). Task is to minimize the average cost  $\overline{C}$ by adapting an optimal decision strategy.  

Defining  {\em risk operator} as,  
\begin{eqnarray} 
R_\alpha=\sum_{\beta}\,  C_{\alpha,\beta}\, {\rm Tr}(\rho_\alpha\, E_\beta),\ \ \end{eqnarray}
one can express the averate cost (\ref{cost}) as, 
\begin{eqnarray} 
\label{risk&cost}
\overline{C}=\sum_{\alpha}\, \Pi_\alpha\,  {\rm Tr}(\rho_\alpha\, R_\beta). \end{eqnarray}
Bayes' strategy~\cite{Bayes58} is to assign the costs     
\begin{equation}
C_{\alpha \beta} =\left\{
\begin{array}{l} 
1,\ {\rm if}\ \alpha\neq \beta,\\ 
0, \ {\rm if}\ \alpha= \beta
 \end{array}
\right. 
\end{equation}
following which the average cost reduces to the minimum {\em average probability of error}:
\begin{eqnarray}
\label{errProb}
P_e&=&\min_{\{E_\beta\}}\, P_{\rm err} = \min_{\{E_\beta\}}\, 
\sum_{\alpha}  \Pi_{\alpha}\,  {\rm Tr}(\rho_{\alpha}\, E_{\beta})   
\end{eqnarray}

Reverting back to the case of binary hypothesis testing, we define the {\em Helstrom matrix}~\cite{Helstrom}:  
\begin{equation}
\Gamma=\Pi_1 \rho_1-\Pi_0 \rho_0.
\end{equation}
Substituting $\displaystyle\sum_{\alpha=0,1}\, E_\alpha=I$, the minimum average probability of error (\ref{errProb}) can be expressed as, 
\begin{equation}\label{errProb2}
P_{e}=\min_{\{E_0, E_1=I-E_0\}}\frac{1}{2}\left\{1+{\rm tr}\left[\Gamma\, (E_0-E_1)\right]\right\}
\end{equation} 
 From the spectral decomposition  of the hermitian Helstrom matrix $\Gamma$,  \begin{equation}
\label{gamma2}
\Gamma= \sum_{k_{+}=1}^{r}\, \lambda_{k_{+}}\vert\,\phi_{k_{+}}\rangle\,\langle\, \phi_{k_{+}}\, \vert 
+\sum_{k_{-}=r+1}^{n}\, \lambda_{k_{-}}\vert\,\phi_{k_{-}}\rangle\,\langle\, \phi_{k_{-}}\, \vert
\end{equation}
in terms of the eigenstates $\vert\,\phi_{k_{\pm}}\rangle, $, corresponding to the real positive/negative eigenvalues $\lambda_{k_{\pm}},$  \ $k_+=1, 2,\ldots r; \, k_-=r+1,r+2,\ldots, n$, we obtain, 
\begin{eqnarray}
	\label{errProb4}
P_{e}&=&\min_{\{E_0, E_1\}}\, \frac{1}{2}\left[1+\left(\sum_{k_{+}=1}^{r}\, \lambda_{k_{+}}\, \langle\,\phi_{k_{+}}\vert\, (E_0-E_1)\vert\,\phi_{k_{+}}\rangle\,
\sum_{k_{-}=r+1}^{n}\, \lambda_{k_{-}}\, \langle\,\phi_{k_{-}}\vert\, (E_0-E_1)\vert\,\phi_{k_{-}}\rangle\,	
	\right)\right]\nonumber \\
\end{eqnarray}
An optimal choice of measurement $\{E_0,\, E_1=I-E_0\}$ turns out to be,  
\begin{eqnarray}
E_0=\sum_{k_{+}=1}^{r}\, \vert\,\phi_{k_{+}}\rangle\,\langle\, \phi_{k_{+}}\vert ,\, \ \  E_1=I-E_0.
\end{eqnarray}
Thus one obtains the {\em minimum} average error probability as 
\begin{eqnarray}
\label{pe}
P_{e} &=& \min_{E_0, E_1}\, P_{\rm err}=\frac{1}{2}\left(1-\vert\vert\,\Gamma\vert\vert\, \right) 
\end{eqnarray}
where $\vert\vert A \vert\vert_1= {\rm Tr}\, \sqrt{A^\dagger\, A}$ denotes the trace norm  of the operator $A$. This result  on {\em single copy minimum probability of error} (given by (\ref{pe})) in testing quantum binary hypotheses is attributed to  Holevo $\&$ Helstrom~\cite{Helstrom,Holevo}.  

 In the symmetric case of equal a priori probabilities, i.e., $\Pi_0=\Pi_1=\frac{1}{2}$, the minimum error probability is given by 
\begin{equation}
\label{perr6}
P_{e}=\frac{1}{2}\left[1-\frac{1}{2}||\rho_1-\rho_0||\right].
\end{equation}
\begin{itemize}
\item If $\rho_0=\rho_1$, then $||\rho_1-\rho_0||=0 \Rightarrow P_e=\frac{1}{2}$, i.e. decision is completely random when the states are identical.  
 \item Minimum probability of error $P_{e}=0$ for orthogonal states $\rho_0$ and $\rho_1$ for which $\vert\vert \rho_0~-~\rho_1\vert=0$ i.e., the states can be discriminated with zero error. 
 \item For  pure states $\rho_0=|\psi_0\rangle\langle\psi_0|$ and $\rho_1=|\psi_1\rangle\langle\psi_1|$, the  
 error probability (\ref{perr6}) gets simplified: 
 \begin{equation}
 \label{perr_pure0}
 P_{e}=\frac{1}{2}\left(1-\sqrt{1-|\langle\psi_0|\psi_1\rangle|^{2}}\right).
 \end{equation}   
\end{itemize}

\subsection{Unambiguous state discrimination}

An {\em unambiguous discrimination} of two quantum states 
with a measurement involving two elements $E_0$, $E_1$ is possible only when the states are orthogonal. In an alternative approach, termed as {\em unambiguous} state discrimination,  introduced by   Ivanovic~\cite{ivanovic_87}, the attempt is to discriminate non-orthogonal states {\em unambiguously} (i.e., with zero error), but the cost that one has to pay in this scheme is due to {\em  inconclusive result} that one ends up with. Here, a POVM consisting of three elements $\{E_0, E_1, E_2=I-E_0-E_1\}$ is chosen. Then, one identifies 
\begin{eqnarray}
\label{3povm1}
{\rm Tr}(\rho_0\, E_1)=0,\ \ {\rm Tr}(\rho_1\, E_0)=0.  
\end{eqnarray} 
But this requires an additional {\em inconclusive result} arising from the measurement element $E_{2}=I-E_0-E_1$ i.e.,  one ends up with uncertainty because  ${\rm Tr}(\rho_0\, E_2)\neq 0$, ${\rm Tr}(\rho_1\, E_2)\neq 0$.  
The errors arising due to inconclusive outcomes are expressed by     
 \begin{equation} 
 \label{3povm2}
{\rm Tr}(\rho_0\, E_2)=1-q_0, \ \  
{\rm Tr}(\rho_1\, E_2)=1-q_1,  \ \ 0\leq q_0,\, q_1 \leq 1
\end{equation}
Using (\ref{3povm1}), and substituting $E_0+E_1+E_2=I$, it  follows that, 
 \begin{eqnarray} 
 \label{unam}
{\rm Tr}(\rho_0\, E_0)&=&{\rm Tr}\,\left(\rho_0\, \{I-E_1-E_2\}\right) =q_0, \nonumber \\
{\rm Tr}(\rho_1\, E_1)&=&{\rm Tr}\,\left(\rho_1\, \{I-E_0-E_2\}\right) =  q_1. 
\end{eqnarray}
Now, consider two pure non-orthogonal states  $\rho_0=\ket{\psi_0}\bra{\psi_0}$, $\rho_1=\ket{\psi_1}\bra{\psi_1}$, occuring with a priori probabilities $\Pi_0,\ \Pi_1$ respectively.  A  measurement scheme with {\em zero discrimination error}, obeying the condition (\ref{3povm1}) can be explicitly  constructed as follows:
\begin{eqnarray}
E_0&=&\frac{q_0}{|\braket{\psi_0|\psi_1^\perp}|^2}\, \vert \psi_1^\perp\rangle \langle \psi_1^\perp\vert  \nonumber \\  
E_1&=& \frac{q_1}{|\braket{\psi_0^\perp|\psi_1}|^2}\, \vert \psi_0^\perp\rangle \langle \psi_0^\perp\vert
\end{eqnarray} 
where $\ket{\psi_0^\perp}$ and 
$\ket{\psi_1^\perp}$ are states orthogonal to $\ket{\psi_0}$ and $\ket{\psi_1}$ respectively. Then we obtain,  
\begin{eqnarray}
\label{pincon}
P_{\rm inconclusive}&=& \Pi_0\, {\rm Tr}\, (\rho_0\, E_2)+\Pi_1\, {\rm Tr}\, (\rho_1\, E_2)   \\
&=& \Pi_0\, q_0+\Pi_1\, q_1 \nonumber 
\end{eqnarray} 
as the {probability of inconclusive result}. 
With the choice  
\begin{eqnarray*} 
q_0 &=& \sqrt{\frac{\Pi_1}{\Pi_0}}\, \vert\langle \psi_0\vert \psi_1\rangle\vert , \\ 
q_1 &=& \sqrt{\frac{\Pi_0}{\Pi_1}}\,  \vert\langle \psi_0\vert \psi_1\rangle\vert .
\end{eqnarray*}
it may be seen that the associated  probability of inconclusive result (\ref{pincon}) reduces to 
\begin{equation}
\label{inc_er}
P_{\rm inconclusive} = 2 \sqrt{\Pi_0\Pi_1}\,\, \vert\langle \psi_0\vert \psi_1\rangle\vert.
\end{equation}
Furthermore, in the symmetric case $\Pi_0=\Pi_1=1/2$,  one ends up with  $P_{\rm inconclusive}=\vert\langle \psi_0\vert \psi_1\rangle\vert $ i.e., the error arising due to inconclusive measurement outcome is proportional to the overlap between the states and is zero only when the states are orthogonal.

\noindent {\bf Comparision of uambiguous state discrimination with Holevo-Helstrom minimum error strategy}\, :  Consider a simple example of discriminating  two non-orthogonal  states $$|\psi_0\rangle=|0 \rangle,\  \ \ |\psi_1\rangle=\frac{|0\rangle +|1\rangle}{\sqrt{2}}.$$ occuring with equal a priori probabilities $\Pi_0=\Pi_1=1/2.$ 
\begin{itemize}
\item The error probability of inconclusive 
outcomes (see (\ref{inc_er}) is given by  $$P_{\rm inconclusive}=1/\sqrt{2}\simeq 0.707.$$ 

\item The minimum probability of error (see (\ref{perr_pure0})) in the Holevo-Helstrom single copy  discrimination scheme is given by,     
 $$P_{e}=\frac{1}{2}\left(1-\sqrt{\frac{1}{2}}\right)\simeq 0.146$$ 
 \end{itemize}
Thus, an experimenter testing which of  the given two states is true one, ends up with  ~70\% error if he/she adapts the unambiguous state discrimination approach. In contrast, using Baysean strategy (which leads to the Holevo-Helstrom result (\ref{perr6}) for discrimination), leads to around  ~15\% error. This example reveals that price to be paid for an {\em error-free} or unambigous  discrimination is high, compared to that for the Baysean minimum error strategy.

\section{Multiple Copy State Discrimination}

Testing hypotheses with multiple copies of quantum states is known to reduce error incurred~\cite{kargin2005, qcb_pra_08, qcb_cmp2008}.  We discuss some known results on error probabilities when $M$ copies of the quantum states,i.e. $\rho_0^{\otimes M}$
and $\rho_1^{\otimes M}$ are available for quantum hypothesis testing. 

Measurement strategies with multiple copies of  quantum states 
are broadly divided into two categories:
\begin{enumerate}
 \item \textbf{Collective measurements:} A single measurement is performed on all the $M$ copies of the quantum states. 
 \item \textbf{Individual measurements:} Each of the measurements (which  may not be the same) are performed separately on individual copies.  
 
\end{enumerate}
We proceed to outline the different measurement strategies.

\subsection{Collective measurements}

The \textit{Holevo-Helstrom} result leading to the error probability (\ref{perr6}) holds in the multiple copy situation too, when an optimal  collective 
measurement is performed on  $\rho_0^{\otimes M}$ and $\rho_1^{\otimes M}$ i.e.,  
\begin{equation}
\label{perr_n}
P_{e}^{(M)}=\frac{1}{2}\left[1-\frac{1}{2}||\rho_1^{\otimes M}-\rho_0^{\otimes M}||_1\right],
\end{equation}
where we have chosen equal a priori probabilities $\Pi_0=\Pi_1=1/2$ for  the states $\rho_0^{\otimes M}$ and $\rho_1^{\otimes M}$ for simplicity.  

We consider some special cases: 
\begin{itemize}
 \item Restricting to pure states $\rho_0=|\psi_0\rangle\langle\psi_0|$ and $\rho_1=|\psi_1\rangle\langle\psi_1|$, the $M$-copy 
 error probability (\ref{perr_n}) reduces to the form,  
\begin{equation}
\label{perr_pure}
P_{e}^{(M)}=\frac{1}{2}[1-\sqrt{1-|\langle\psi_0|\psi_1\rangle|^{2 M}}]
\end{equation}
 As $0<|\langle\psi_0|\psi_1\rangle|<1$  the error probability (\ref{perr_pure}) declines 
by increasing the number of copies  $M$. For $M>>1$ and  $\langle\psi_0|\psi_1\rangle^{2M}<<1$, we obtain, 
\begin{eqnarray}
\label{approx}
P_{e}^{(M)}&\approx& \frac{1}{2}\left[1-\left(1-\frac{1}{2}|\langle\psi_0|\psi_1\rangle|^{2 M}\right)\right]= \frac{1}{4} |\langle\psi_0|\psi_1\rangle|^{2 M} \label{perr_pure_M}.
\end{eqnarray} 

\item In the aysmptotic limit of $M\rightarrow \infty$, the $M$-copy error probability declines exponentially~\cite{qcb_prl_07,qcb_cmp2008, qcb_pra_08}  
\begin{equation}
\label{perr_expo}
P_{e}^{(M)}\sim e^{-M\,  \xi_{\rm QCB}}\ \ \  {\rm as}\ \ M\!\to\!\infty.
\end{equation} 
where the optimal error exponent $\xi_{\rm QCB}(\rho_0,\rho_1)$ is given by, 
\begin{equation}
\label{chernoff_information}
\xi_{\rm QCB}=\inf\limits_{s\in[0,1]} {\rm log \ Tr}\{ \rho_0^s \rho_1^{1-s}\}.
\end{equation}

\item An upper bound on the $M$ copy error probability  $P_{e}^{(M)}$ of (\ref{perr_n})  has been established~\cite{qcb_prl_07, qcb_pra_08, qcb_cmp2008},
\begin{equation}
\label{qcb_ub}
P_{e}^{(M)}\leq P_{QCB}^{(M)},
\end{equation}  
 based on the
quantum Chernoff error exponent $\xi_{\rm QCB}$, where the error upper bound given by  
\begin{equation}
\label{qcb_def}
P_{QCB}^{(M)}=\frac{1}{2}(\inf_{0\leq s \leq 1} {\rm Tr}\{\rho_0^s \rho_1^{1-s}\})^M
\end{equation}
is referred to as the quantum Chernoff Bound (QCB).
\end{itemize}

\subsection{Individual measurements}

It is known that  collective measurements perform better than separeate measurements done  on individual copies of the $M$-copy state resulting in  optimal state discrimination when multiple copies of the states $\rho_0$, $\rho_1$ are given~\cite{Peres&Wooters91}. But, when the number of copies $M$ is large, collective measurements are hard to implement experimentally. Thus it is of interest to explore how far one may be able to approach results of optimal state discrimination (realized based on collective measurement strategy)  by confining to individual measurements i.e., to measurements performed separately on each copy of the collective $M$-copy states $\rho_0^{\otimes M}$, $\rho_1^{\otimes M}$.     

\noindent {\bf Fixed individual measurements}: Consider $\rho_0^{\otimes M}$ and $\rho_1^{\otimes M}$
occuring equal probabilitiies $\Pi_0=1/2$ and $\Pi_1=1/2$. Consider a individual measurement scheme, where same measurement is performed {\em individually} on each copy of 
$\rho_0^{\otimes M}$, $\rho_1^{\otimes M}$. In the specific case with measurements  $E_0^{\otimes M}$ and $E_1^{\otimes M}$, with individual measurement operators  $E_0=\ket{\psi_0}\bra{\psi_0}$ and $E_1=I-\ket{\psi_0}\bra{\psi_0}$,  the error probability  $P_{\rm ind}^{(M)}$ is given by,
\begin{equation}
\label{peind1}
P_{\rm ind}^{(M)} =\frac{1}{2}\, \left( {\rm Tr}[\, \rho_0^{\otimes M} E_1^{\otimes M}\, ]+  {\rm Tr}[\, \rho_1^{\otimes M} E_0^{\otimes M}\, ]\right). 
\end{equation}  
\begin{itemize}
\item If one of the states, say $\rho_0$ is pure, i.e. $\rho_0^{\otimes M}=\left(|\psi_0\rangle \langle \psi_0\vert\right)^{\otimes M}$, 
the error probability (\ref{peind1}) can be simplified: 
\begin{eqnarray}
P_{\rm ind}^{(M)} &=& 
\frac{1}{2}\, \left( {\rm Tr}[\rho_0^{\otimes M}\, E_1^{\otimes M}]+{\rm Tr}[\rho_1^{\otimes M}\, E_0^{\otimes M}] \right) \nonumber \\ 
&=&\frac{1}{2}\left( {\rm Tr}[\rho_0^{\otimes M}\,(I-|\psi_0^{\otimes M}\rangle \langle \psi_0^{\otimes M}|)]+{\rm Tr}[\,( \rho_1^{\otimes M}\,|\psi_0^{\otimes M}\rangle \langle \psi_0^{\otimes M}|)\,] \right)\nonumber \\
\label{perr_ind}
&=&\frac{1}{2} \langle \psi_0|\rho_1|\psi_0\rangle^{M}.  
\end{eqnarray}

\item If both the states are pure, then the error probability simplifies to 
\begin{equation}
\label{pe_local_pure1}
P_{\rm ind}^{(M)}=\frac{1}{2} |\langle \psi_0|\psi_1\rangle|^{2M}.
\end{equation}
Note that the approximate value $P_{e}^{(M)}\approx |\langle \psi_0|\psi_1\rangle|^{2M}/4$  of $M$-copy error probability realized using collective measurement strategy ( see (\ref{approx})) is less than $P_{\rm ind}^{(M)}$. In other words, error probability obtained using collective  measurements provides a lower bound on that realized from individual fixed measurements. They both match (i.e., they approach the value $0$) only in the asymptotic limit $M\rightarrow \infty$.    

\end{itemize}

\noindent {\bf Adaptive Measurements} : In an adaptive measurement scheme, the restriction on fixed measurement on each copy of the state is relaxed. The strategy here is  to optimize the next consequent measurement by using the information gathered from the results of previous measurement. This is done in a step by step manner. It has been shown~\cite{acin_pra_2005} that local adaptive measurements can reveal equally good performance as that of collective optimized measurements. Further details about  adaptive measurement strategy can be found in  references~\cite{acin_pra_2005, adaptive_non_apadtive_pra_2010, global_local_pra_11,  wiseman_17}.

\noindent {\bf Bounds on error probability}: 
Recall that quantum  fidelity $F(\rho_0,\rho_1)$ defined by~\cite{ulhman, josza, nc2000}  
\begin{equation}
\label{fidelity_def}
F(\rho_0,\rho_1)=\left[{\rm Tr}\left(\sqrt{\sqrt{\rho_0}\rho_1\sqrt{\rho_0}}\right)\right]^2,
\end{equation} 
serves as a quantitative measure of how close are the states $\rho_0$ and $\rho_1$.   
It is known that the trace norm  $||\rho_1-\rho_0||_1$  is bounded by the fidelity  $F(\rho_0,\rho_1)$ as follows~\cite{nc2000}:
\begin{equation}
\label{fidelity_tn}
1-\sqrt{F(\rho_0,\rho_1)}\leq \frac{1}{2}||\rho_1-\rho_0||_1 \leq \sqrt{1-F(\rho_0,\rho_1)}.
\end{equation} 
Using the property $F\left(\rho_0^{\otimes M},\, \rho_1^{\otimes M} \right)=
\left[F\left(\rho_0,\, \rho_1 \right)\right]^M$,  the following upper and lower bounds~\cite{kargin2005} are realized on the optimal $M$-copy error probability (see (\ref{perr6})): 
\begin{equation}
\label{fidelity_bounds}
\frac{1}{2}\, \left(1-\sqrt{1-[F(\rho_0,\rho_1)]^{M}}\right)\leq P_{e}^{(M)}\leq \frac{1}{2}\, \left(\sqrt{F(\rho_0,\rho_1)}\right)^{M}.
\end{equation} 
\begin{itemize}
\item If one of the states is pure, say  $\rho_0=\ket{\psi_0}\bra{\psi_0}$, a strict upper bound on optimal error probability  $P_{e}^{(M)}$ follows: 
\begin{equation}
\label{fidelity_strict}
 P_{e}^{(M)}\leq \,\frac{1}{2} |\langle \psi_0|\psi_1\rangle|^{M}.
\end{equation} 
\item When both the states are pure i.e.,  $\rho_0=\ket{\psi_0}\bra{\psi_0}$ and  $\rho_1=\ket{\psi_1}\bra{\psi_1}$, the lower bound in (\ref{fidelity_bounds}) matches with the exact expression (\ref{perr_pure}) on $M$-copy error probability. 
\end{itemize}

Another pair of computable  upper and lower bounds, referred to as quantum Bhattacharya bounds~\cite{qcb_pra_08, OgaHay}  are found to be useful in identifying the asymptotic limit of the $M$-copy error probability: 
(\ref{perr6}) 
\begin{equation}
\label{bh_bounds}
\frac{1}{2}\, \left(1-\sqrt{1-\left[{\rm Tr}\left(\rho_0^{\frac{1}{2}}\,\rho_1^{\frac{1}{2}}\right)\right]^{2M}}\right)\leq P_{e}^{(M)}\leq \frac{1}{2}\, \left[{\rm Tr}\left(\rho_0^{\frac{1}{2}}\,\rho_1^{\frac{1}{2}}\right)\right]^{M}.
\end{equation} 

\begin{itemize}
\item The upper bounds of (\ref{fidelity_bounds}) and (\ref{bh_bounds})  are related to each other as, 
\begin{eqnarray}
F(\rho_0,\rho_1)=\left({\rm Tr}[\sqrt{\sqrt{\rho_0}\rho_1\sqrt{\rho_0}}]\right)^2&=& \left({\rm Tr}[\sqrt{\sqrt{\rho_0}\sqrt{\rho_1}\sqrt{\rho_1}\sqrt{\rho_0}}]\right)^2\nonumber \\
&=& \vert\vert\rho_0^{\frac{1}{2}}\,\rho_1^{\frac{1}{2}}\vert\vert_1^2,
\end{eqnarray}
  leading to~\cite{qcb_pra_08}
\begin{equation}
{\rm Tr}[\sqrt{\rho_0}\sqrt{\rho_1}]\leq  \vert\vert\, \sqrt{\rho_1}\sqrt{\rho_0}\, \vert\vert_1\equiv 
\sqrt{F(\rho_0,\rho_1)}. 
\end{equation}
Thus, one obtains the inequality constraining the $M$-copy error probability: 
\begin{equation}
P_{e}^{(M)}\leq \frac{1}{2}\, \left[{\rm Tr}\left(\rho_0^{\frac{1}{2}}\,\rho_1^{\frac{1}{2}}\right)\right]^{M}\leq \frac{1}{2}\, \left(\sqrt{F(\rho_0,\rho_1)}\,\right)^{M}
\end{equation} 
\end{itemize}

\section{Quantum Channel Discrimination}

Suppose that an input  quantum state $\rho$ goes through channels $\Phi_{\alpha}, \ \alpha=0,1,2,\ldots$. The channels $\Phi_{\alpha}$ acting on $\rho$ result in the output states  $\rho_\alpha$ of the channel: 
\begin{equation}
\Phi_\alpha(\rho) = \rho_{\alpha}.
\end{equation} 
The task is to ascertain which of the channels $\{\Phi_\alpha\}$ the state $\rho$ went through. We confine  here to discrimination of two channels $\Phi_0$, $\Phi_\alpha$. 

The question of distinguishing  channels $\Phi_\alpha, \alpha=0,1$ by choosing an input state $\rho$ reduces to that of detecting  the output  states $\rho_0$ and $\rho_1$  (see Fig.~1) with an appropriate measurement strategy.  The single copy error-probability for binary channel discrimination is given by   
\begin{eqnarray}
\label{channel_perr}
P_{e}&=&\frac{1}{2}\,\left(1-\frac{1}{2}\vert\vert\Phi_0(\rho)-\Phi_1(\rho)\vert\vert_1\,\right)
\nonumber\\
&=& \frac{1}{2}\,\left(1-\frac{1}{2}\vert\vert\rho_0-\rho_1\vert\vert_1\,\right).
\end{eqnarray}

\begin{figure}
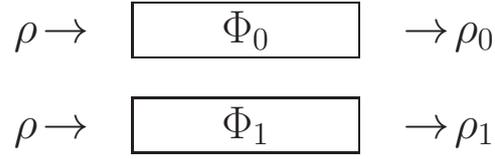

{\Large
\vspace{-1.5cm}
\begin{align}
\rho \!\to\! \hspace{0.5cm} \framebox[3cm][c]{$\Phi_0$}\hspace{0.5cm} \!\to\! \rho_0 \nonumber \\
\rho \!\to\! \hspace{0.5cm} \framebox[3cm][c]{$\Phi_1$}\hspace{0.5cm} \!\to\! \rho_1 \nonumber
\end{align}
}
\caption{Discrimination of quantum channels $\Phi_0,\ \Phi_1$}
\end{figure}
An optimization over a set of all input states $\rho$ leads to the minimum error probability of discriminating the two channels $\Phi_0$, $\Phi_1$ i.e., 
\begin{align}
\label{channel_perr_2}
\min\limits_{\rho\in \mathcal{H}} P_{e}
 &= \frac{1}{2}\left(1-\max\limits_{\{\rho\in \mathcal{H}\}}\,\frac{1}{2}\,||\Phi_0(\rho) - \Phi_1(\rho)||_1\right).
\end{align}

\subsection{Entanglement as a resource for channel discrimination}

Consider a composite bipartite state $\rho_{AB}\in \mathcal{H}_A\otimes \mathcal{H}_B$ as  an input to the channel(s) $\Phi_0$ ($\Phi_1$). The channels $\Phi_0$, $\Phi_1$ are designed so as to act only on one of  the subsystems, say $\rho_A={\rm Tr}_B(\rho_{AB})$.    
It is convenient to employ the notation  $\Phi_0=\Phi_0^{(A)}$ and $\Phi_1=\Phi_1^{(A)}$. Action of the channels on the input state $\rho_{AB}$ is expressed as follows:
\begin{align}
[\Phi_0\otimes \mathbbm{1}] (\rho_{AB}) &=\rho_{AB}^{(0)} \nonumber \\ 
[\Phi_1\otimes \mathbbm{1}] (\rho_{AB}) &=\rho_{AB}^{(1)}. \nonumber
\end{align}
Here $\mathbbm{1}$ denotes the identity channel. 

The error probability $P_{\rm e}$ of discriminating binary channels, in a single evaluation, is given by, 
\begin{align}
\label{channel_perr_ent}
P_{e} &=\frac{1}{2}\left(1-\frac{1}{2}||\rho^{(0)}_{AB} - \rho^{(1)}_{AB}||\right) \nonumber \\
&=\frac{1}{2}\left(1-\frac{1}{2}\left|\left|[\Phi_0\otimes \mathbbm{1}] (\rho_{AB}) - [\Phi_1\otimes \mathbbm{1}] (\rho_{AB})\right|\right|\right).
\end{align}
and the minimum error probability is obtained by optimizing over the set of all input  bipartite states $\rho_{AB}$, 
\begin{align}
\label{channel_perr_ent_2}
\underset{\left\{\rho_{AB}\in\mathcal{H}_A\otimes\mathcal{H}_B\right\}}{\min}\, P_{e} &=\frac{1}{2}\left(1- \max\limits_{\left\{\rho_{AB}\in\mathcal{H}_A\otimes\mathcal{H}_B\right\}}\,\frac{1}{2} \,\left|\left|[\Phi_0^{(A)}\otimes \mathcal{I}^{(B)}] (\rho_{AB}) - [\Phi_1^{(A)}\otimes \mathcal{I}^{(B)}] (\rho_{AB})\right|\right|\right)  \\
 &=\frac{1}{2}\left(1-\frac{1}{2} ||\Phi_0-\Phi_1||_{\diamond}\right)
\end{align}  
where  $$||\Phi_0-\Phi_1||_{\diamond}=\max\limits_{\rho_{AB}\in\mathcal{H}_A\otimes\mathcal{H}_B}\left|\left|[\Phi_0^{(A)}\otimes \mathcal{I}^{(B)}] (\rho_{AB}) - [\Phi_1^{(A)}\otimes \mathcal{I}^{(B)}] (\rho_{AB})\right|\right|$$ 
is referred to as the {\em diamond} norm~\cite{kitaev_97, watrous08}. 

An optimization over the set of all pure bipartite states is enough~\cite{opt_dis_sacchi_pra_2005, watrous_piani_09} for achieving minimum error probability in \eqref{channel_perr_ent_2}.  In the case of single-shot channel discrimination, Piani and Watrous~\cite{watrous_piani_09} have shown that  
\begin{align}
\max\limits_{\rho^{(\rm sep)}_{AB}\in\mathcal{H}_A\otimes \mathcal{H}_B)}\, \left|\left|[\Phi_0^{(A)}\otimes \mathcal{I}^{(B)}] (\rho_{AB}) - [\Phi_1^{(A)}\otimes \mathcal{I}^{(B)}] (\rho_{AB})\right|\right|
=\max\limits_{\rho\in \mathcal{H_A}}\, ||\Phi_0^{(A)}(\rho) - \Phi_1^{(B)}(\rho)||,
\end{align}
 Here an optimization is carried out by restricting only to the set of all separable states $$\rho^{(\rm sep)}_{AB}=\sum_{i}\, p_i\, \rho_{A,i}\otimes  \rho_{B,i}, \ 0\leq p_i\leq 1,\ \sum_i\, p_i=1.$$  
 In other words, there is {\em no advantage} in employing a {\em separable} composite bipartite state $\rho^{(\rm sep)}_{AB}$ as  input of the channels, because the probability of error does not get reduced beyond the one achievable using any input state $\rho$ belonging to the Hilbert space  $\mathcal{H}_A$ itself.  On the otherhand, it has been identified that entangled input states help    in channel  discrimination   with improved
 precision~\cite{opt_dis_sacchi_pra_2005,ent_breaking_sacchi_pra_05,Lloyd2008,Lloyd_prl,qtd_aru_09,our_qr_13,which_path_pra_14}, where it has been established that with a choice of entangled input state, it is possible to reduce channel discrimination error probability. More specifically, Piani and Watrous~\cite{watrous_piani_09} proved,    
\begin{equation}
\label{diamond_eq}
||\Phi_0-\Phi_1||_{\diamond} \geq \max\limits_{\rho^{(\rm sep)}_{AB}\in\mathcal{H}_A\otimes \mathcal{H}_B)}\,\left|\left|[\Phi_0^{(A)}\otimes \mathcal{I}^{(B)}] (\rho_{AB}) - [\Phi_1^{(A)}\otimes \mathcal{I}^{(B)}] (\rho_{AB})\right|\right|.
\end{equation}
in the case of single-shot discrimination of the channels. In other words, given an entangled state,  it is always possible to find a pair of quantum channels such that the  error probability of single-shot channel discrimination gets minimized. For a detailed  mathematical treatement on channel discrimination see Ref.~\cite{watrous08}.

\noindent {\bf Discrimination of identity and completely depolarizing channels}: Let us consider an example\cite{opt_dis_sacchi_pra_2005, qtd_aru_09}, where a completely depolarizing channel  and an identity channel, labeled respectively 
as channel 0 and channel 1,  are to be discriminated based on their action on  a pure  input state $\vert\psi\rangle$, belonging to a finite $d$ dimensional Hilbert space ${\cal H}_d$.  The output states of the channels are given by,  
\begin{eqnarray}
\rho_0&=&\Phi_0(\rho)=\frac{I}{d} \nonumber \\ 
\rho_1&=&\Phi_1(\rho)=\vert\psi\rangle\langle\psi\vert.
\end{eqnarray}
The single copy error probability in distinguishing $\rho_0$ and $\rho_1$ is readily found to be, 
\begin{eqnarray}
P_{e,\vert\psi\rangle}&=&\frac{1}{2}\,\left(1-\frac{1}{2}\left\vert\left\vert \frac{I}{d}-
\vert\psi\rangle\langle\psi\vert\, \right\vert\right\vert_1\,\right)\nonumber \\
&=& \frac{1}{2}\,\left(1-\frac{1}{2}\left[\left\vert \frac{1}{d}-1\right\vert + 
\frac{d-1}{d}\right]\right)=\frac{1}{2d}.
\end{eqnarray}
Let us consider a   maximally entangled $d\times d$  state, 
\begin{equation}
\label{me}
\vert\Psi_{AB}\rangle=\frac{1}{\sqrt{d}}\sum_{i=1}^{d}\vert i_A,i_B\rangle,
\end{equation}
as input of the channels $\Phi_0\otimes \mathbbm{1}$, $\Phi_0\otimes \mathbbm{1}$. The output states are then found to be,   
\begin{eqnarray}
\rho_{AB}^{(0)}&=&(\Phi_0\otimes \mathbbm{1})\, \vert\Psi_{AB}\rangle\langle\Psi_{AB}\vert\, \nonumber \\
&=&\frac{I}{d}\otimes{\rm Tr}_A[\vert\Psi_{AB}\rangle\langle\Psi_{AB}\vert\,]=\frac{I\otimes I}{d^2}\nonumber \\
\rho_{AB}^{(1)}&=&(\Phi_1\otimes \mathbbm{1})\,\vert\Psi_{AB}\rangle\langle\Psi_{AB}\vert=\vert\Psi_{AB}\rangle\langle
\Psi_{AB}\vert.
\end{eqnarray}
The error-probability in discriminating the two channels, with a maximally entangled state is equal to  
\begin{equation}
P_{e,\vert\Psi_{AB}\rangle}=\frac{1}{2d^2}.
\end{equation}
This clearly shows that maximally entangled state (\ref{me}) is advantageous in the discrimination of completely depolarizing and identity channels~\cite{opt_dis_sacchi_pra_2005, qtd_aru_09}. 

\section{Summary}

This article presents an overview of quantum state discrimination based on binary hypothesis testing. A brief outline on {\em Unambiguous state discrimination}, an alternate approach developed for quantum state discrimination, is given, with the help of an illustrative example. Collective and adaptive measurements strategies employed in the case of multiple copy hypothesis testing are described. A discussion  on computable upper and lower bounds on error probability  in the multiple copy scenario and  the error rate exponent in the asymptotic limit is given. Furthermore,  quantum  channel descrimination and the role of entangled states in enhancing precision in the task of channel discrimination are presented.

\section*{acknowledgements} 
ARU and SRA are supported by the University Grants Commission (UGC) Major Research Project
(Grant No. MRP-MAJOR-PHYS-2013-29318), Govern-
ment of India. JPT acknowledges support from UGC-BSR, India.

\bibliography{qht}

\end{document}